# Auxetic Black Phosphorus: A 2D Material with Negative Poisson's Ratio

Yuchen Du[1,3], Jesse Maassen[1,3,4,*], Wangran Wu[1,3], Zhe Luo[2,3], Xianfan Xu[2,3,*], and Peide D. Ye[1,3,*]

[1] School of Electrical and Computer Engineering, Purdue University, West Lafayette, Indiana 47907, United States

[2] School of Mechanical Engineering, Purdue University, West Lafayette, Indiana 47907, United States

[3] Birck Nanotechnology Center, Purdue University, West Lafayette, Indiana 47907, United States

[4] Department of Physics and Atmospheric Science, Dalhousie University, Halifax, Nova Scotia, Canada, B3H 4R2

* Address correspondence to: yep@purdue.edu (P.D.Y.); xxu@ecn.purdue.edu (X.X.); jmaassen@dal.ca (J.M.)

ABSTRACT

The Poisson's ratio of a material characterizes its response to uniaxial strain. Materials normally possess a positive Poisson's ratio - they contract laterally when stretched, and expand laterally when compressed. A negative Poisson's ratio is theoretically permissible but has not, with few exceptions of man-made bulk structures, been experimentally observed in any natural materials. Here, we show that the negative Poisson's ratio exists in the low-dimensional natural material black phosphorus, and that our experimental observations are consistent with first principles simulations. Through application of uniaxial strain along zigzag and armchair directions, we find that both interlayer and intralayer negative Poisson's ratios can be obtained in black phosphorus. The phenomenon originates from the puckered structure of its in-plane lattice, together with coupled hinge-like bonding configurations.

MAIN TEXT

When a material is stretched in one direction by Δ$L$, as depicted in Figure 1a, it usually tends to contract in the other two directions perpendicular to the direction of stretching. Similarly, when a material experiences a compressive force, it expands laterally in the directions perpendicular to the direction of compression[1,2]. In both cases, the magnitude of deformation is governed by one of the fundamental mechanical properties of materials, the so-called Poisson's ratio. The Poisson's ratio of a material defines the ratio of the transverse contraction to the longitudinal extension in the direction of the stretching force. Specifically, it quantitatively explains how much a material becomes thinner (or thicker) in lateral directions when it experiences a longitudinal tension (or compression)[1,2]. For ordinary materials, the Poisson's ratio is always positive. However, the possibility that the Poisson's ratio becomes negative has been an accepted concept in the classical elasticity theory for over 160 years[1,3], implying that, a material with a negative Poisson's ratio would undergo a transverse contraction when compressed, and a transverse expansion when stretched in the longitudinal direction. Although a negative Poisson's ratio is theoretically permitted, direct observation of such a phenomenon in natural materials has never happened. Materials with a negative Poisson's ratio, also named auxetic materials, were demonstrated in 1987 by Lakes in a designed re-entrant (bow-tie) form[4]. Since then, auxetic materials have been extensively studied in macroscopic bulk form with microscopically engineered structures, as they can be useful in medicine, tissue engineering, bulletproof vest, and fortified armor enhancement. The negative Poisson's ratio in these man-made structures is derived from controlling the geometry and deformation mechanism of the internal material structure from the macroscopic level down to the molecular level[1,5-10]. While the concept of auxetic materials with special artificial microstructures has gradually become accepted within the past decades, the question remains whether a negative Poisson's ratio exists in natural materials. In fact, theoretical predictions of a negative Poisson's ratio in naturally occurring single crystal was proposed back in the early 70s[11,12], and also performed on a variety of crystal models[13-17]. However, there is a lack of experimental evidence since the measurement of internal deformation in auxetic materials, in particular at the atomic level, is extremely difficult. In practice, the in-plane Poisson's ratio measurement was conducted by recording the movements of location markers on a surface during constant-rate deformation, and the cross-plane Poisson's ratio was obtained from utilizing scanning electron microscopy to generate vertical distance variations with applied in-plane strains[18]. Even though previous efforts have been carried out in studying the auxetic behavior among man-made materials and structures, we are not aware of any reports on the experimental demonstration of the negative Poisson's ratio in naturally occurring crystals at atomic structure accuracy as we present here on a 2D auxetic material – black phosphorus (BP).

BP, a stable phosphorus allotrope at room temperature[19], is a layered natural semiconducting crystal composed of sheets of monolayer phosphorene, and therefore it can be mechanically exfoliated into atomically thin layers with a vertical dimension of a couple of nanometers down to one monolayer[20-28]. The importance of exploring isolated thin-film BP is built on the fact that it bridges the gap between zero bandgap graphene and wide bandgap transition metal dichalcogenides, thereby providing a new route to expand the scope of experimentally accessible 2D crystals, and pursue a broad range of the fundamental studies. BP exhibits a thickness-dependent bandgap characteristic, ranging from ~0.3 eV in bulk crystal to >1.4 eV in the form of a monolayer[20-25]. Its p-type nature and high carrier mobility are also valuable contributions to the family of 2D materials. The moderate bandgap, along with relatively high carrier mobility, also benefits BP in electronic and optoelectronic applications[20-28]. Here, we focus on another property of BP: its unique puckered structure allows BP to exhibit substantial anisotropy in the mechanical properties with respect to strains[29-38], and encourages us to experimentally and theoretically investigate and demonstrate the existence of negative Poisson's ratio.

In this study, we employed Raman spectroscopy to experimentally demonstrate the cross-plane interlayer (between adjacent monolayer phosphorene layers) negative Poisson's ratio when it is uniaxially strained along the armchair direction. Furthermore, we confirmed the existence of the cross-plane intralayer (within a monolayer phosphorene layer) negative Poisson's ratio under uniaxial deformation along the zigzag direction. In contrast to man-made bulk structures, these multiple negative Poisson's ratios are intrinsic to BP, and they are attributed to the puckered structure along its in-plane anisotropic axes, where the unique lattice structure can be regarded as a natural re-entrant form that is comprised of two coupled hinge-like bonding configurations [35].

**Results**

**Lattice vibration modes and polarized Raman characterization of BP.** In a unit cell of BP, each phosphorus atom covalently bonds to its three nearest neighbors, forming warped hexagons. As shown in Figure 1b, this $sp^3$-type bonding has introduced a distinctly anisotropic crystal structure resulting in two principal lattice axes referred to as armchair ($x$-direction) and zigzag ($y$-direction), which are perpendicular and parallel to the pucker, respectively[20]. The geometric anisotropy introduced by the pucker implies that BP would exhibit significant anisotropic lattice vibration response to uniaxial strain along armchair or zigzag directions[35,36], which can be observed directly from Raman spectroscopy[37,38]. Previous studies of Raman spectra have shown that there are three prominent active modes in BP[22,23]. The cross-plane $A_g^1$ mode occurs due to opposing vibrations of top and bottom phosphorus atoms with respect to each other. The $B_{2g}$ mode describes the bond movement along the in-plane zigzag direction. The $A_g^2$ mode has a dominate component along the in-plane armchair direction, as illustrated in Figure 1c. To start our experiment, few-layer BP was exfoliated from the bulk crystal by the standard scotch tape method and transferred to a conducting Si substrate with a 300 nm $SiO_2$ capping layer. Polarized Raman spectroscopy was utilized to determine the flake orientation[28]. With the detection polarization parallel to the incident laser polarization, the active phonon mode of $B_{2g}$ is not detected due to matrix cancellation when the two principle lattice axes are aligned with the laser polarization[28]. The $A_g^2/A_g^1$ Raman intensity ratio can further be used to distinguish the specific armchair or zigzag axis. The intensity of the armchair-oriented $A_g^2$ mode is maximized and is about twice the intensity of the $A_g^1$ mode when the laser polarization is along the armchair direction. The intensity of $A_g^2$ is comparable to $A_g^1$ when laser is aligned along the zigzag direction[28]. The candidate BP flakes were all pre-characterized by the polarized Raman system. Optical and atomic force microscope (AFM) images of a representative 7.3 nm thick BP flake are presented in Figures 1e and 1f, respectively. The armchair and zigzag lattice axes in Figure 1e were determined by the $A_g^2/A_g^1$ intensity ratio shown in Figure 1d.

**The interlayer negative Poisson's ratio in armchair strained BP.** We first investigate the evolution of the Raman spectra of BP with uniaxial tensile and compressive strains, summarized in Figure 2a. The laser polarization is aligned along the zigzag axis of BP flake, and the strain direction is along armchair direction based on our apparatus set-up (see Supplementary note 1). In our work, we used Lorentzian functions to fit the Raman spectra and obtained the peak frequency of each mode at different strains. For unstrained BP, consistent with previous reports[22,23], we observe the cross-plane vibration mode of $A_g^1$ at ~362 cm$^{-1}$, and in-plane vibration modes of $B_{2g}$ and $A_g^2$ at ~439 cm$^{-1}$ and ~467 cm$^{-1}$, respectively. The $A_g^1$ and $B_{2g}$ modes show the same linear trend of Raman frequency shift with respect to the applied strain, while the rate of frequency shift is different in these two modes. Both $A_g^1$ and $B_{2g}$ modes experience a red-shift when BP is under tensile strain along the armchair direction, with a slope of 1.37 cm$^{-1}$%$^{-1}$ and 1.07 cm$^{-1}$%$^{-1}$, respectively. On the other hand, $A_g^1$ and $B_{2g}$ have a blue-shift at a rate of 1.78 cm$^{-1}$%$^{-1}$ and 0.88 cm$^{-1}$%$^{-1}$ under uniaxial compressive strain, as shown in Figures 2b and 2c. It is worth mentioning that we did not observe a measurable Raman shift in the $A_g^2$ mode, corresponding to the lateral vibration in the armchair direction, with armchair strains (see Supplementary note 2), and we believe this can be attributed to the fact that the BP structure is anisotropic and is much softer along the armchair direction, compared to the zigzag direction. The sensitivity of determining strains in BP using Raman peak positions is greater than that of molybdenum disulfide (MoS$_2$), but slightly smaller than those of carbon nanotubes and graphene[39,40]. The error bars in the figures are extracted from the Lorentzian peak fittings, which are significantly smaller than the strain-induced frequency shift. For the applied strain less than 0.2 %, the Raman peak position remains the same at each strain level after multiple loading and unloading cycles, indicating our experiments are highly reliable (see Supplementary note 3). In addition, the absence of discrete jumps of any of the three Raman modes under monotonically varied strains assures that the BP flake does not slip against the substrate during strain experiments.

The different responses of Raman spectra with strains can be explained by analyzing the types of vibration mode involved[36]. Let us take the $A_g^1$ mode as an example, where the atomic motions occur mainly along the cross-plane direction. As depicted in the inset of Figure 2b, the $A_g^1$ mode vibration in few-layer BP is determined by two components, the interlayer distance $I_Z$ and the intralayer phosphorus bond length $d_I$. The red-shift of the $A_g^1$ mode under tensile strain along the armchair direction, with a slope of 1.37 cm$^{-1}$%$^{-1}$, can be understood on the basis of the elongation of the P-P intralayer bond length $d_I$, which weakens the interatomic interactions and therefore reduces the vibration frequency. Also, there exists another possibility that the red-shifted Raman frequency of the $A_g^1$ mode is attributed to the enlarged interlayer distance $I_Z$, where reduced interlayer interactions can also weaken the cross-plane vibration frequency[41,42]. If this is true, then it is interesting that BP expands (both intralayer bond length $d_I$ and interlayer distance $I_Z$) when it is stretched along the armchair direction, which is opposite to the definition of positive Poisson's ratio. Meanwhile, the blue-shift of 1.78 cm$^{-1}$%$^{-1}$ under compressive armchair strain indicates the intralayer bond length and interlayer distance may be smaller under compressive armchair strain, thus proposing a hypothesis that the cross-plane Poisson's ratio is negative in BP with an armchair full strain.

To corroborate our experimental findings that suggest a negative Poisson's ratio in the cross-plane direction, and to determine the individual contributions from $I_Z$ and $d_I$, we performed density functional theory (DFT) calculations of strained BP (details are described in the Methods section). Since the experimental thickness of the BP samples is ~7 nm, for our modeling purposes we focus on bulk BP. Bulk BP is characterized by three lattice constants $a$, $b$ and $c$, aligned along the armchair, zigzag, and cross-plane directions, respectively (see Figure 3a). Figure 3b shows how the lattice constants are modified with uniaxial strains along the armchair directions. Interestingly, we also observe a negative Poisson's ratio from our DFT calculations: +1 % tension (or -1 % compression) of the armchair lattice parameter, $a$, enhances +0.5 % (or reduces -0.4 %) the cross-plane lattice parameter, $c$. In other words, BP becomes "thicker" in the vertical direction as it undergoes armchair tensile strain, and it becomes "thinner" in the vertical direction under armchair compressive strain. This supports the hypothesis derived from the experimental observation that the cross-plane Poisson's ratio is negative for the entire armchair strain. The value of the Poisson's ratio, defined as $\nu \approx -\frac{\Delta L'}{\Delta L}$, where $\Delta L'$ and $\Delta L$ are the variation in the cross plane direction and in the armchair direction, respectively are thus -0.5 under tension and -0.4 under compression. Note that the magnitude of the strain values assumed in the calculations is equal or larger than 1% to ensure accurate results well above the numerical error.

To understand the origin of how the cross-plane lattice parameter $c$ is modified with strain, in Figure 3c we present how the various atomic distances, and their projections, are modified due to armchair strain. The lattice parameter $c$ is controlled by both the $z$-projection of the cross-plane P-P bond $d_{1Z}$ and the interlayer distance $I_Z$. The intralayer $d_{1Z}$ shows a regular positive Poisson's ratio, meaning the constituent monolayer phosphorene layers flatten (or expand) when stretched (or compressed) laterally along the armchair direction. This is consistent with previous calculations of armchair strain in monolayer phosphorene[35]. However, the interlayer $I_Z$ shows an unusual negative Poisson's ratio behavior, where the distance between the monolayer phosphorene layers expands by +0.9 % (or shrinks by -0.7 %) when stretched (or compressed) laterally along the armchair direction by 1%. More importantly, the change in $I_Z$ under armchair strain is more pronounced than that of $d_{1Z}$, meaning the cross-plane lattice parameter $c$ is dominated by the interlayer distance $I_Z$ rather than $d_{1Z}$. Thus the negative Poisson's ratio associated with $c$ originates from the interlayer coupling in BP, and we call this phenomenon interlayer negative Poisson's ratio, which is the first such demonstration in a naturally occurring 2D material. This also suggests that the interlayer interaction is not purely van der Waals in BP, but may also arise from the coupling between the wavefunctions of lone pair electrons in adjacent layers[42-44]. Naturally this result was not predicted from the previous phosphorene studies[35], which only considers a single layer, but did find a negative Poisson's ratio for $d_{1Z}$ with zigzag strain (to be discussed in the following section). Within a small magnitude of strain, the first-order negative Poisson's ratio approximation yields, $\nu \approx -\frac{\Delta L'}{\Delta L}$ where $\Delta L$ and $\Delta L'$ are lattice constant variation and bond length variation with respect to armchair lattice and interlayer distance $I_Z$. Specifically, BP demonstrates an interlayer negative Poisson's ratio of -0.9 along stretching armchair, and -0.7 under armchair compression from DFT calculations.

To connect the strain induced atomic structural change to that of the Raman response, we calculated the phonons frequencies of bulk BP under strain. In Figure 1c, we show the atomic displacements associated with the three Raman-active phonon modes, $A_g^1$, $B_{2g}$ and $A_g^2$. As noted above the $A_g^1$ mode corresponds to atomic motion predominantly in the cross-plane direction, and is thus likely to be most sensitive to variations in cross-plane distances (*i.e.* $d_{1Z}$ and $I_Z$). Figure 3d presents the variation in phonon frequency versus armchair strain. Focusing on the $A_g^1$ mode, we find that the frequency increases with increasing armchair compressive strain. From a microscopic point of view, this can be understood as follows: when BP is compressed along armchair, the main effect is to modify the $d_1$ bond angle such that it aligns more with the cross-plane direction (in Figure 3c we see a large change in $d_{1X}$ with little change to $d_1$). In this process the $d_1$ bond becomes more closely aligned with the atomic displacement of the $A_g^1$ mode, which leads to an increase in $A_g^1$ frequency. With tensile armchair strain, the $d_1$ bond becomes less aligned with the atomic displacements of the $A_g^1$ mode and the energy decreases. This effect of bond angle alignment influences both the P atoms within a single layer (intralayer) and the nearest P atoms in adjacent layers (interlayer). Note that, interestingly, as the atoms in adjacent layers are brought closer through a modification of the $d_1$ bond angle via compressive armchair strain $I_Z$ decreases, which enhances interlayer vibration and also increase $A_g^1$ frequency. While we find reasonable agreement with $A_g^1$, there is some discrepancy for the $B_{2g}$ and $A_g^2$ modes. For example, we observe an opposite change in $B_{2g}$ phonon energy with strain compared to experiment. We note, however, that the strain effect is relatively small. We believe that this difference could arise from the interaction of BP with the substrate, which is not captured in the calculations, or due to the larger adopted strain values in the DFT modeling (>1% theory versus <0.2% experiment). Nevertheless, the calculated changes in atomic structure due to strain, which are typically more robust than the more sensitive changes in phonon energy, show a negative Poisson's ratio. Thus overall our theoretical results on the effect of strain on the atomic structure and the phonon energies are in good agreement with experimental data, supporting the existence of interlayer negative Poisson's ratio in layered BP along armchair uniaxial strain. This property is significantly different from other 2D materials, *i.e.*, $MoS_2$[38,45].

**The intralayer negative Poisson's ratio in zigzag strained BP.** To apply strain along the zigzag direction, the same BP sample substrate was rotated by 90°, and the polarized Raman spectra was applied again to verify the BP orientation. The incident polarized laser light is aligned along the armchair direction. The corresponding Raman spectra of zigzag strained BP are presented in Figure 4a. Figure 4b illustrates the $A_g^1$ peak position as a function of tensile and compressive zigzag strain. The dominate $A_g^1$ mode peak overlaps with another peak appearing as a weak shoulder. In this case, we fitted the data by two Lorentzian functions and treated the main peak as the peak for the $A_g^1$ mode. This phenomenon only happens when the incident polarized light is along the armchair lattice of BP flakes, and it has been observed in our previous studies as well[28]. The $A_g^1$ mode shows a monotonic behavior for the full range of zigzag strain with a slope of 0.56 cm$^{-1}$%$^{-1}$ under tensile strain and 0.86 cm$^{-1}$%$^{-1}$ under compressive strain. On the other hand, we observe a much more pronounced $B_{2g}$ and $A_g^2$ sensitivity with respect to the zigzag strain as compared to the armchair strain case. Specifically, $B_{2g}$ sensitivities under tensile (5.46 cm$^{-1}$%$^{-1}$) and compressive (5.27 cm$^{-1}$%$^{-1}$) zigzag strains are 5-6 times larger than those under armchair strains. Meanwhile, the peaks of the $A_g^2$ mode also shift under zigzag strains, with about 2.73 cm$^{-1}$%$^{-1}$ and 2.22 cm$^{-1}$%$^{-1}$ for tensile and compressive strains, respectively. This giant anisotropic strain response in BP is directly associated with its anisotropic lattice structure, where the armchair axis is much softer with a smaller Young's modulus as compared to the zigzag axis. A previous theoretical study has predicted a negative Poisson's ratio in monolayer phosphorene along the cross-plane direction under uniaxial zigzag deformation[35]. Here, our DFT simulations of bulk BP also show a negative cross-plane intralayer Poisson's ratio for the zigzag strained BP, summarized in Figure 5a. Figure 5c shows that the calculated thickness of the constituent monolayer phosphorene layers, $d_{IZ}$, increases by +0.1 % with 1 % zigzag stretching and decreases by -0.1 % with 1 % zigzag compression, demonstrating an intralayer negative Poisson's ratio. Therefore, if we consider the intralayer deformation only, its Poisson's ratio is approximately -0.1 under zigzag strain, which is consistent with the previous report on monolayer phosphorene[35]. However, the interlayer distance $I_Z$ shows a positive Poisson ratio characteristic (see Figure 5c) which counteracts the effect of $d_{IZ}$, thus leading to an overall positive Poisson's ratio for the lattice constant $c$ with zigzag strain (see Figure 5b). The calculated variations in the $B_{2g}$ and $A_g^2$ phonon frequency due to zigzag strain are given in

Figure 5d. The sensitivity of the $B_{2g}$ mode vs. strain is nearly twice as high as that of the $A_g^2$ mode, agreeing well with the experimental results.

**Conclusion**

In summary, we have investigated the anisotropic strain responses of few-layer BP films under uniaxial tensile and compressive strains. For the first time, by examining the Raman evolution of uniaxially strained BP, we have succeeded in demonstrating a cross-plane negative Poisson's ratio when it is strained along the armchair direction. *Ab* initio calculations have also been carried out to determine the influence of strain on the atomic structure of BP as well as on the phonon modes. Our theoretical results are consistent with experiment, and indicate there exists an interlayer negative Poisson's ratio between the phosphorene layers under armchair strain. Meanwhile, our results support the existence of a cross-plane intralayer negative Poisson's ratio in the constituent phosphorene layers under uniaxial deformation along the zigzag axis, which is in line with a previous theoretical prediction. In contrast to man-made auxetic materials, this is the first time a negative Poisson's ratio is observed experimentally in a natural material and confirmed by theoretical simulations.

**Methods**

**Sample preparation.** Few-layer films were exfoliated from the bulk crystal BP (Smart-elements), and then transferred to a 525 μm thick Si substrate with 300 nm $SiO_2$ capping layer. The thickness of the BP was measured using a Veeco Dimension 3100 AFM system. During the Raman measurement, the BP flake was exposed to the air for about 2 hours, but with no significant degradation (see Supplementary note 4) in terms of lattice vibration modes.

**Raman measurements.** All Raman measurements were carried out on a HORIBA LabRAM HR800 Raman spectrometer. The system is equipped with a He-Ne excitation laser (wavelength 632.8 nm), an 1800 gmm$^{-1}$ grating, and a Nikon ×50 (NA = 0.45) long-working-distance objective lens. For polarized Raman characterization of the BP flakes, a linear polarizer (Thorlabs, LPNIRE050-B) was used as the analyzer. Subsequent Raman spectroscopy of strained BP studies was performed under an excitation laser power of 0.17 mW, sufficiently low to avoid excessive sample heating.

**First principles calculations.** DFT calculations were carried out using the Vienna *Ab*-initio Simulation Package[46,47] (VASP), based on a plane-wave expansion of the wavefunctions and the Projector Augmented Wave[48] (PAW) method to treat the effect of the core. Our calculations used the generalized gradient approximation (GGA) within the PBE approach for exchange-correlation potential. A plane-wave cutoff energy of 750 eV, and a Monkhorst-Pack *k*-grid of 17×23×23 were adopted to ensure proper convergence of the total energy (<10 μeV/atom). The simulation cell is defined by the following lattice vectors: $a_1$ = [$a$ 0 0], $a_2$ = [0 $b$/2 −$c$/2], $a_3$ = [0 $b$/2 $c$/2], where the lattice constants are determined to be $a$ = 4.564 Å, b = 3.305 Å and c = 11.318 Å (in the case of no strain). In order to include the effect of uniaxial strain the lattice constant along the direction of strain was fixed, and the other two lattice constants were optimized to minimize the total energy (at each step the atoms were relaxed until the forces were < 0.001 eV/Å). The Γ-point phonon mode energies/frequencies were calculated within the harmonic approximation using the finite displacement method to extract the 2$^{nd}$ order force constants, as implemented in the Phonopy software package[49]. The force constants were obtained from 3×4×4 supercell self-consistent calculations using a 4×4×4 Γ-centered *k*-grid.

FIGURES

**Figure 1.**

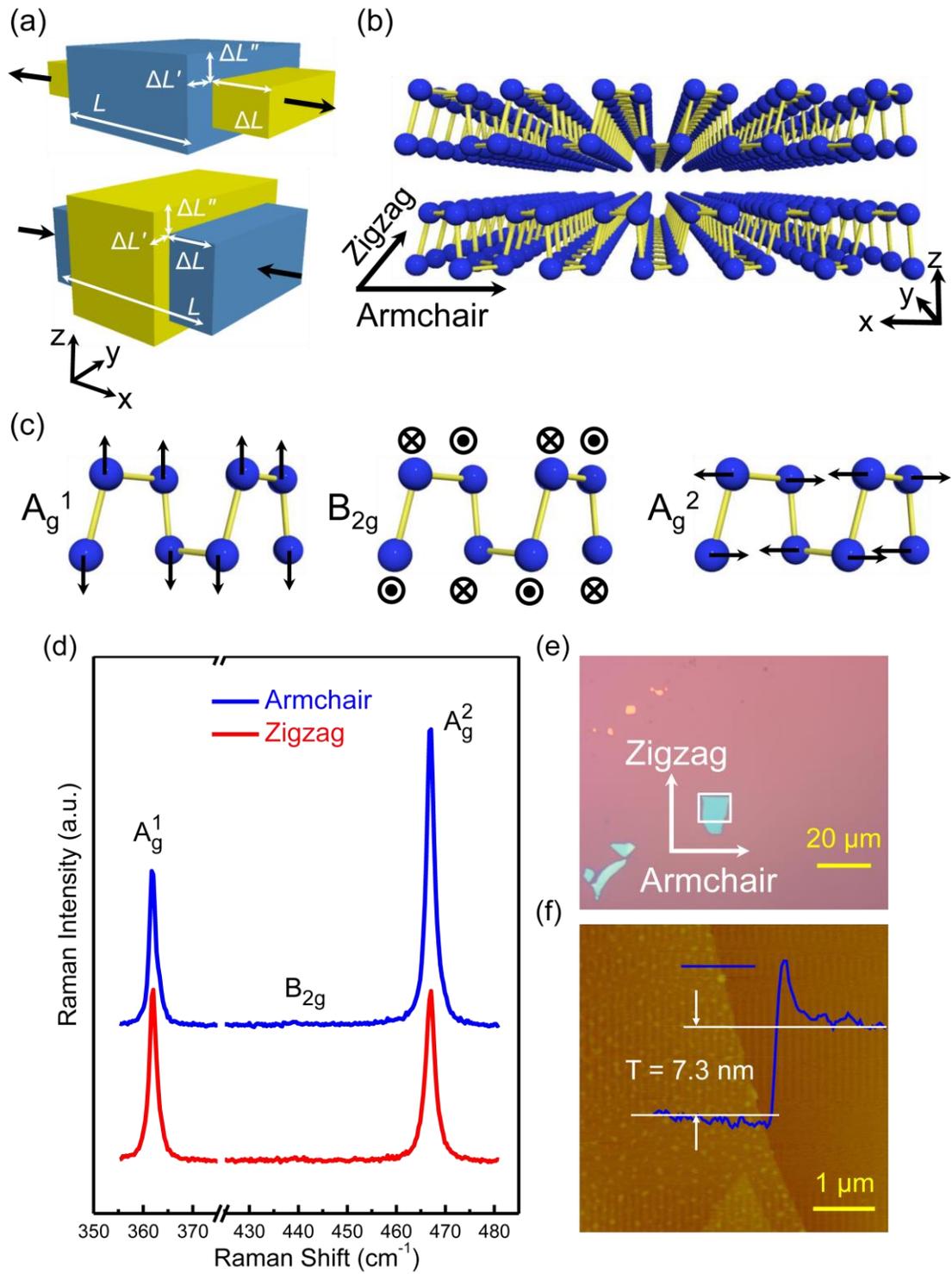

**Figure 2.**

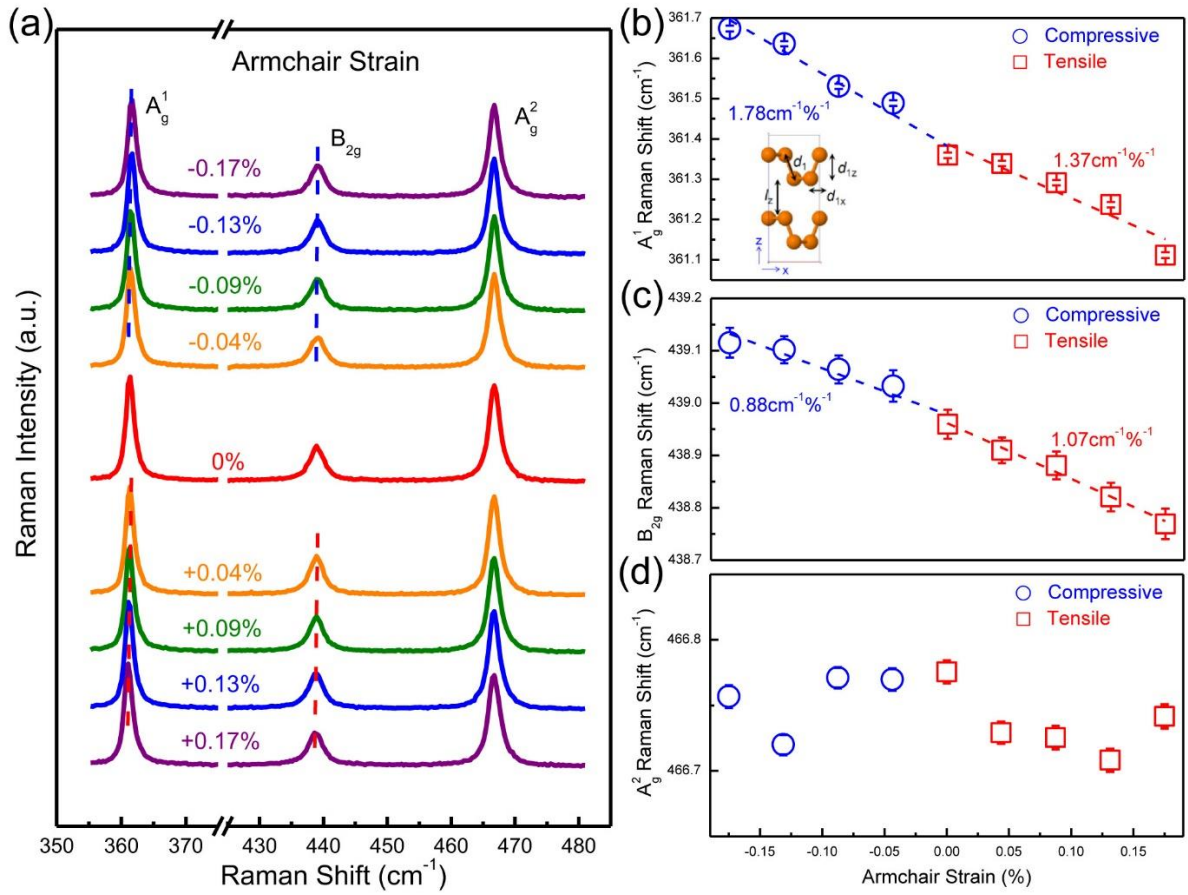

**Figure 3.**

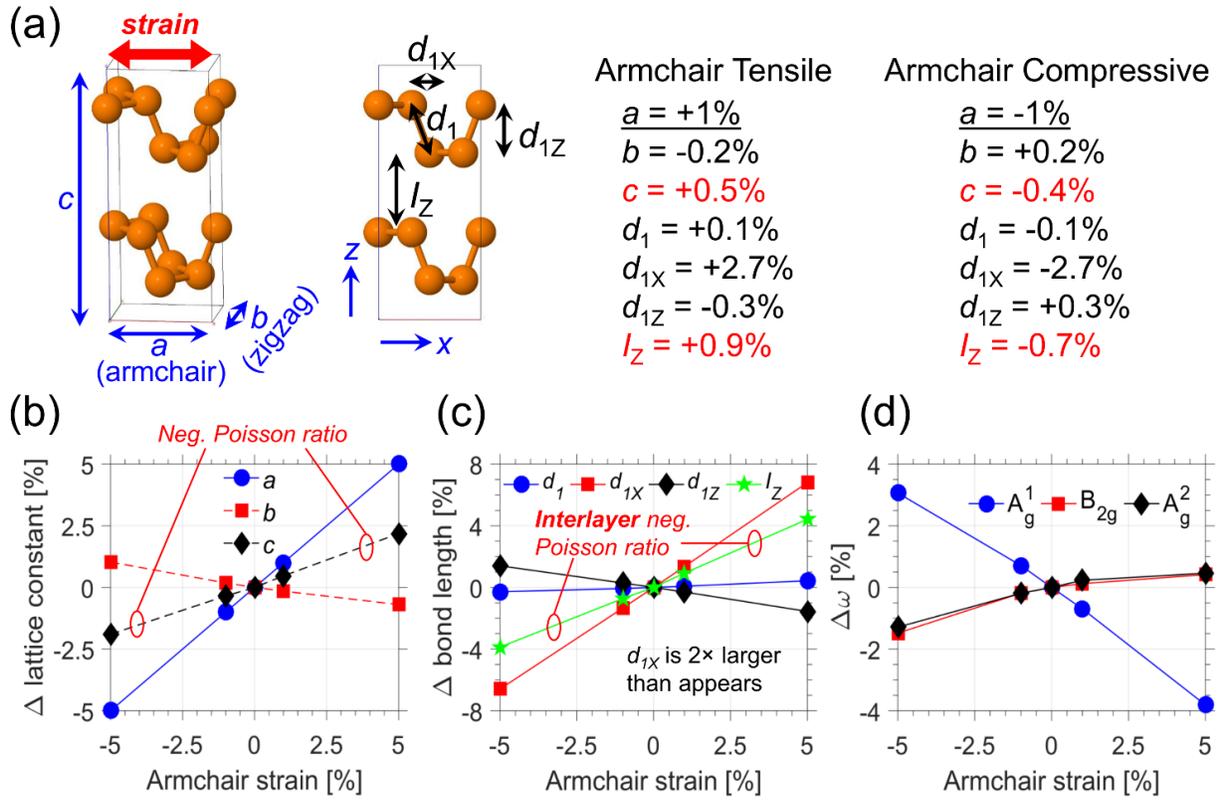

**Figure 4.**

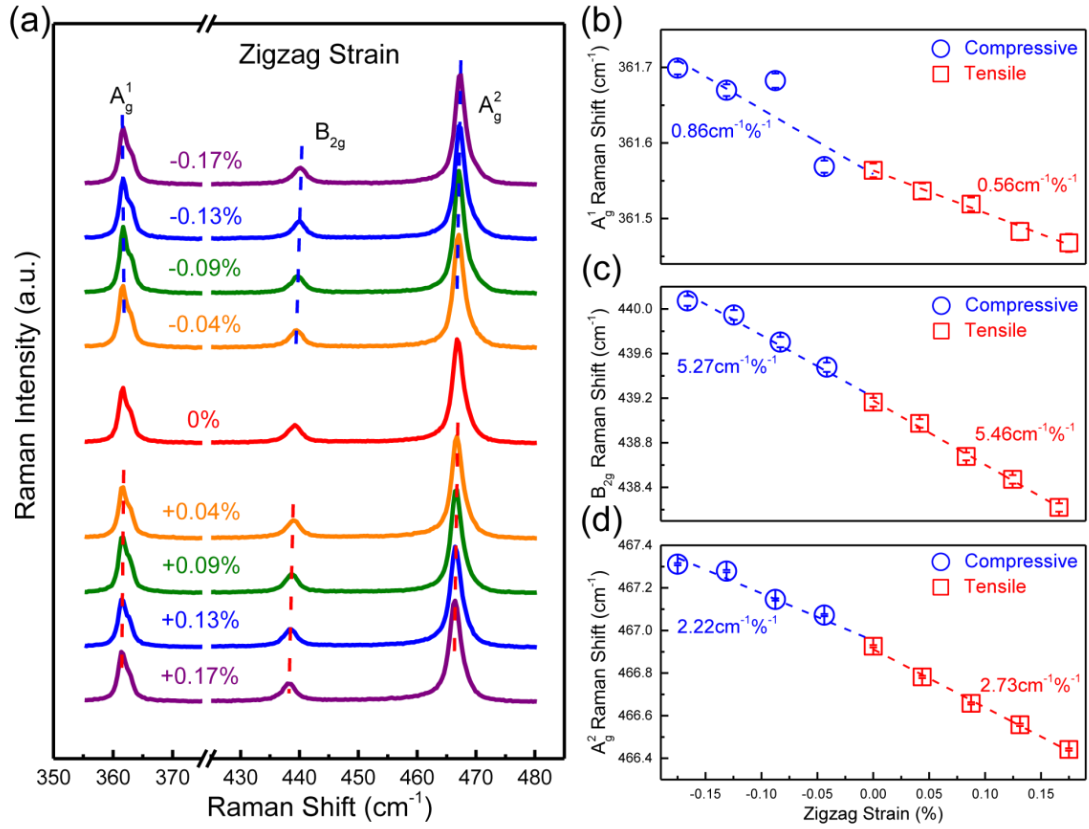

**Figure 5.**

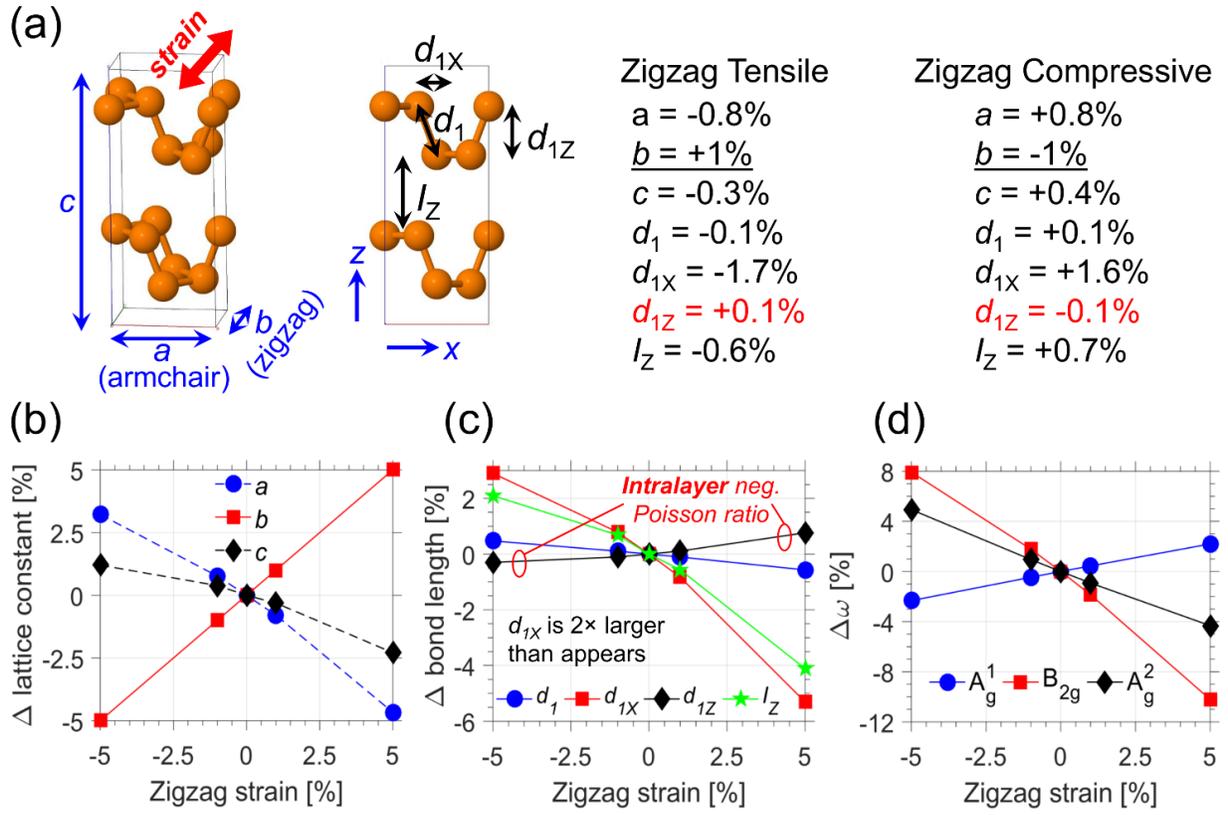

**Figure 1 | BP characterization.** (a) Schematic view of the positive Poisson's effect. A cube with sides of length $L$ of an anisotropic linearly elastic material subject to both tensile and compressive strains along $x$ axis. The blue box is unstrained. The yellow box is stretched (compressed) in the $x$-direction by $\Delta L$, and contacted (expanded) in the $y$ and $z$-directions by $\Delta L'$ and $\Delta L''$, respectively. (b) Lattice structure of BP and (c) atomic vibrational patterns of $A_g^1$, $B_{2g}$ and $A_g^2$ phonon modes. (d) Polarized Raman spectra of BP to distinguish the armchair and zigzag axes. Armchair direction has $A_g^2/A_g^1$ intensity ratio of ~2, while zigzag direction has a ratio of ~1. (e) Optical image of the 7.3 nm thick BP flake with two principle lattice orientations. The white box indicates the laser focusing location. Scale bar in optical image is 20 μm. (f) AFM topology of measured BP flake, and the scale bar is 1 μm.

**Figure 2 | Raman evolution of uniaxial armchair strained BP.** (a) Raman spectra of BP for both tensile and compressive armchair strains. The dashed lines are here to guide the Raman peak position shift. Raman shift of (b) $A_g^1$, (c) $B_{2g}$, (d) $A_g^2$ modes in armchair strained BP. The dashed lines show linear fit results, and error bars are determined from Lorentzian peak fittings.

**Figure 3 | Simulation of armchair strained BP.** (a) Atomic structure of bulk BP. The lattice constants along the main crystals directions are indicated with blue arrows. The cross-plane bond ($d_1$), along with its $x$ and $z$ projections, and the interlayer distance ($I_z$) are indicated with black arrows. (b) Calculated variation in lattice constants due to uniaxial strain along the armchair direction. The solid line indicates the lattice parameter that is fixed to apply the strain. (c) Calculated variation in bond lengths due to uniaxial strain along the armchair direction. The $d_{1X}$ projected bond length is twice as large as appears, to more easily visualize the data in this figure. (d) Calculated variation in the $A_g^1$, $B_{2g}$ and $A_g^2$ phonon frequency due to uniaxial strain along the

armchair direction. A positive strain value corresponds to stretching the BP lattice along the armchair direction, and a negative strain value corresponds to compressing the lattice.

**Figure 4 | Raman evolution of uniaxial zigzag strained BP.** (a) Raman spectra of BP for both tensile and compressive zigzag strains. The dashed lines are here to guide the Raman peak position shift. Raman shift of (b) $A_g^1$ (c) $B_{2g}$ and (d) $A_g^2$ modes in zigzag strained BP. The dashed lines show linear fit results, and error bars are determined from Lorentzian peak fitting.

**Figure 5 | Simulation of zigzag strained BP.** (a) Atomic structure of bulk BP. The lattice constants along the main crystals directions are indicated with blue arrows. The cross-plane bond ($d_1$), along with its $x$ and $z$ projections, and the interlayer distance ($I_z$) are indicated with black arrows. (b) Calculated variation in lattice constants due to uniaxial strain along the zigzag direction. The solid line indicates the lattice parameter that is fixed to apply the strain. (c) Calculated variation in bond lengths due to uniaxial strain along the zigzag direction. The $d_{1X}$ projected bond length is twice as large as appears, to more easily visualize the data in this figure. (d) Calculated variation in the $A_g^1$, $B_{2g}$ and $A_g^2$ phonon frequency due to uniaxial strain along the zigzag direction. A positive strain value corresponds to stretching the BP lattice along the zigzag direction, and a negative strain value corresponds to compressing the lattice.

ACKNOWLEDGEMENTS

This material is based upon work partly supported by NSF under Grant ECCS-1449270, AFOSR/NSF under EFRI 2-DARE Grant EFMA-1433459, and ARO under Gant W911NF-14-1-0572. J.M. acknowledges financial support from NSERC of Canada. The authors would like to thank Mark S. Lundstrom for valuable discussions.

AUTHOR CONTRIBUTIONS

P.D.Y. and X.X. conceived the idea, designed and supervised the experiments. Y.D. and W.W. performed the strain experiments and analyzed the experimental data. J.M. conducted the theoretical calculations and analyses. Y.D., W.W., Z.L. and X.X. performed the polarized Raman experiments and analyses. Y.D., J.M., W.W., Z.L., X.X. and P.D.Y. co-wrote the manuscript.

COMPETING FINANCIAL INTERESTS STATEMENT

The authors declare no competing financial interests.

Supplementary Information for:

# Auxetic Black Phosphorus: A 2D Material with Negative Poisson's Ratio

Yuchen Du[1,3], Jesse Maassen[1,3,4,*], Wangran Wu[1,3], Zhe Luo[2,3], Xianfan Xu[2,3,*], and Peide D. Ye[1,3,*]

[1] School of Electrical and Computer Engineering, Purdue University, West Lafayette, Indiana 47907, United States

[2] School of Mechanical Engineering, Purdue University, West Lafayette, Indiana 47907, United States

[3] Birck Nanotechnology Center, Purdue University, West Lafayette, Indiana 47907, United States

[4] Department of Physics and Atmospheric Science, Dalhousie University, Halifax, Nova Scotia, Canada, B3H 4R2

\* Address correspondence to: yep@purdue.edu (P.D.Y.); xxu@ecn.purdue.edu (X.X.); jmaassen@dal.ca (J.M.)

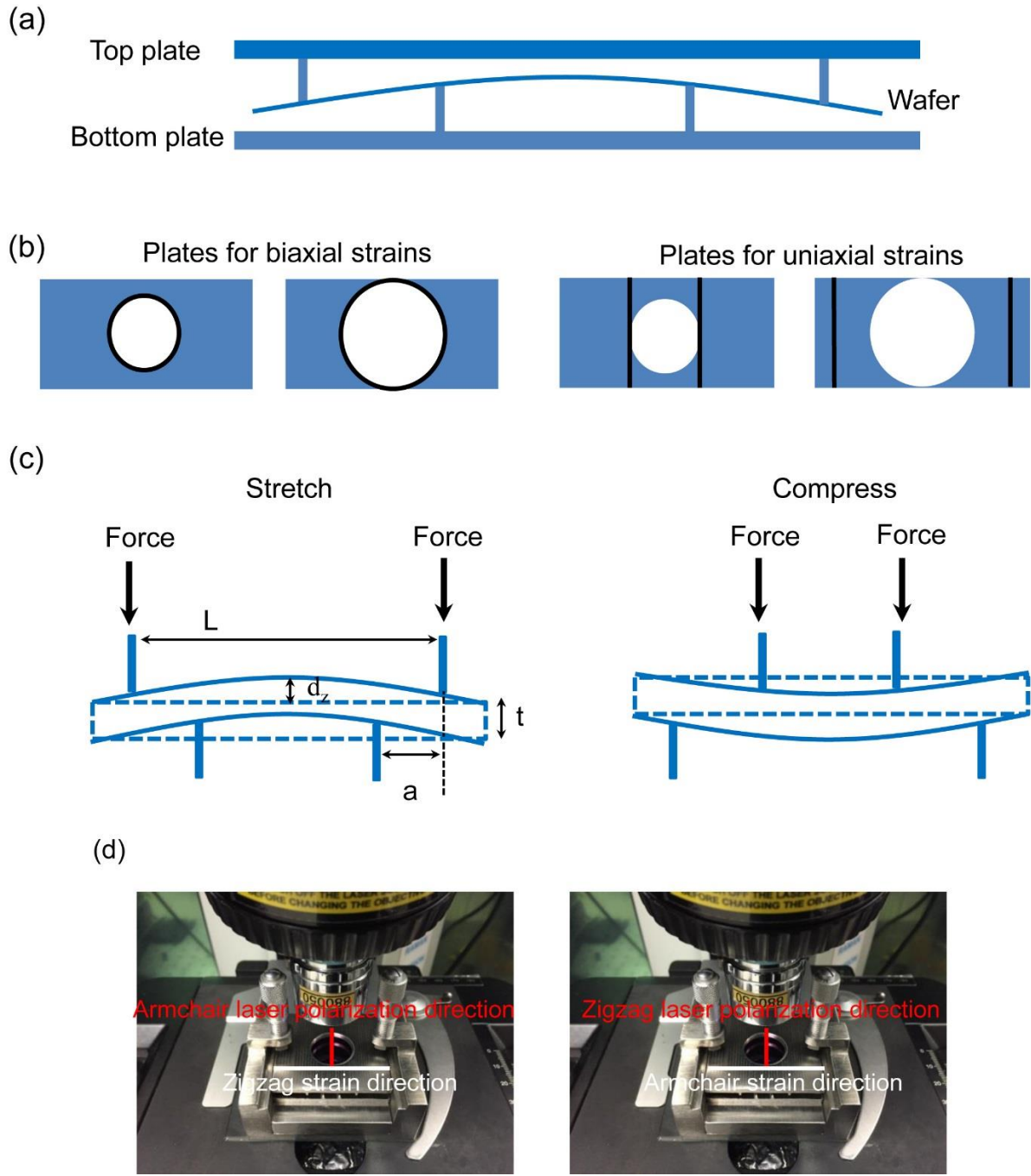

**Supplementary Figure 1 | Schematic structure of uniaxial four-point bending apparatus.** (a) Side-view of a four-point bending apparatus structure under a uniaxial tensile process. (b) Top-view of two plates for biaxial and uniaxial strains. The black lines indicate the metal legs as shown in (a). (c) Schematic views of both tensile and compressive operations. (d) Optical image of strain apparatus set-up. The red line indicates laser polarization direction, and the white line stands for applied strain direction.

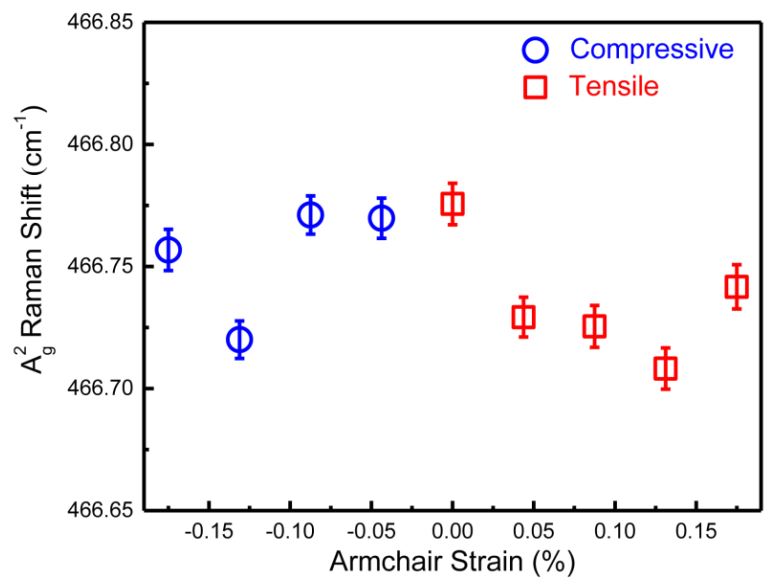

**Supplementary Figure 2 | $A_g^2$ Raman shift in armchair strained BP.** Insensitive $A_g^2$ peak in armchair strained BP for both tensile and compressive processes.

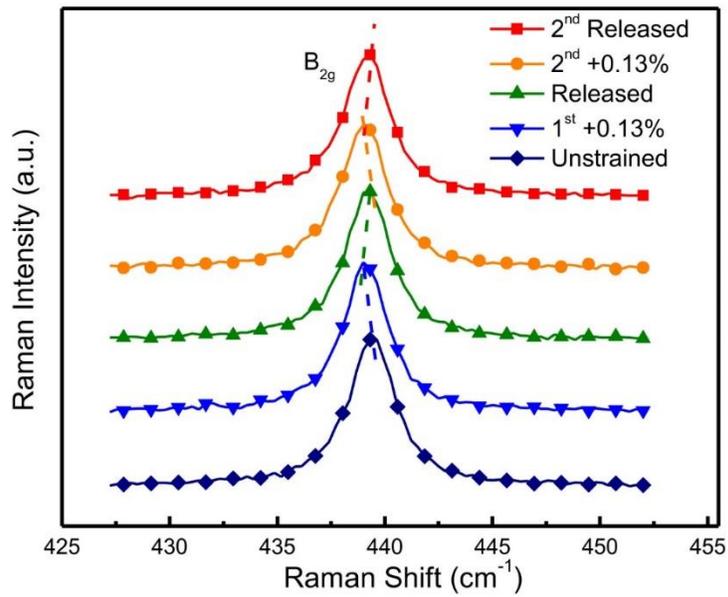

**Supplementary Figure 3 | Multiple loading and unloading processes from Raman spectroscopy measurement.** Raman spectra of uniaxial strained BP for multiple loading and unloading processes.

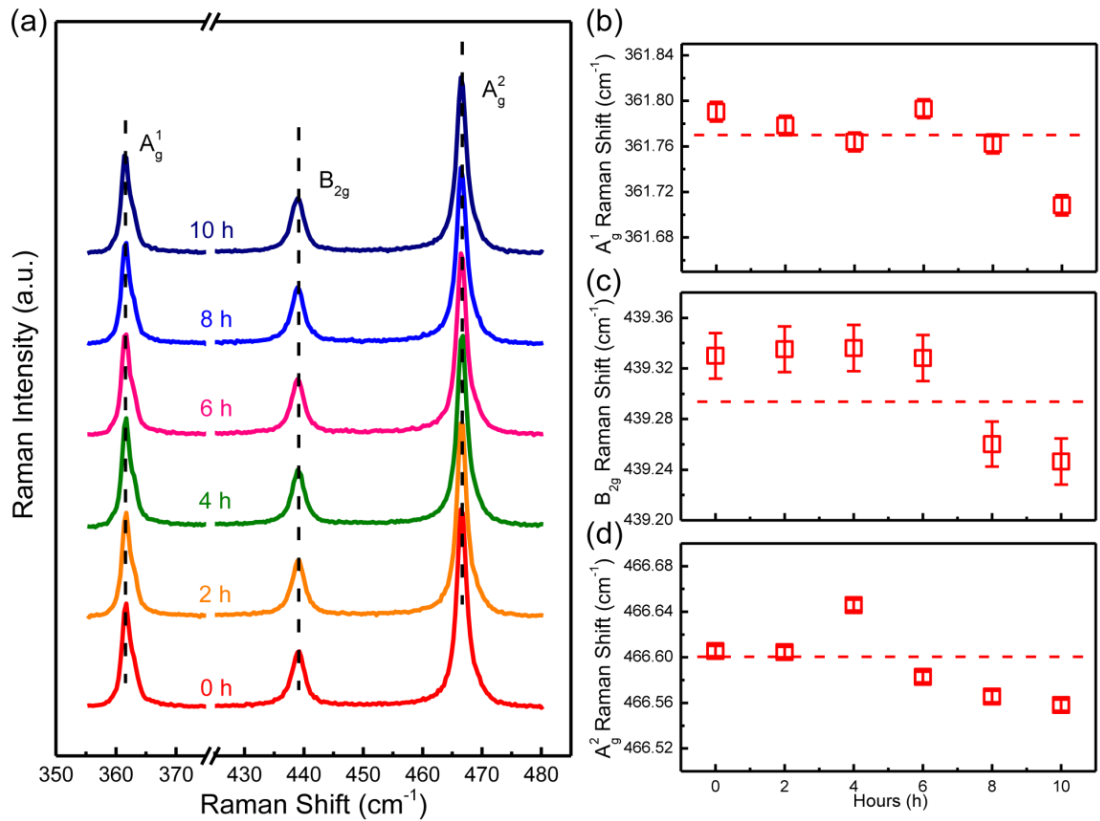

**Supplementary Figure 4 | Air-stability studies of BP Raman spectroscopy measurement.** (a) The evolution of Raman intensity as a function of time up to 10 hours. The peak positions of (b) $A_g^1$ (c) $B_{2g}$, and (d) $A_g^2$ modes for various time. The error bars are determined from Lorentzian curve fitting. The dashed lines in (b), (c) and (d) are the average values in each mode. The variation in (b)-(d) within 10 hours is much less than the strained induced systematic Raman shift.

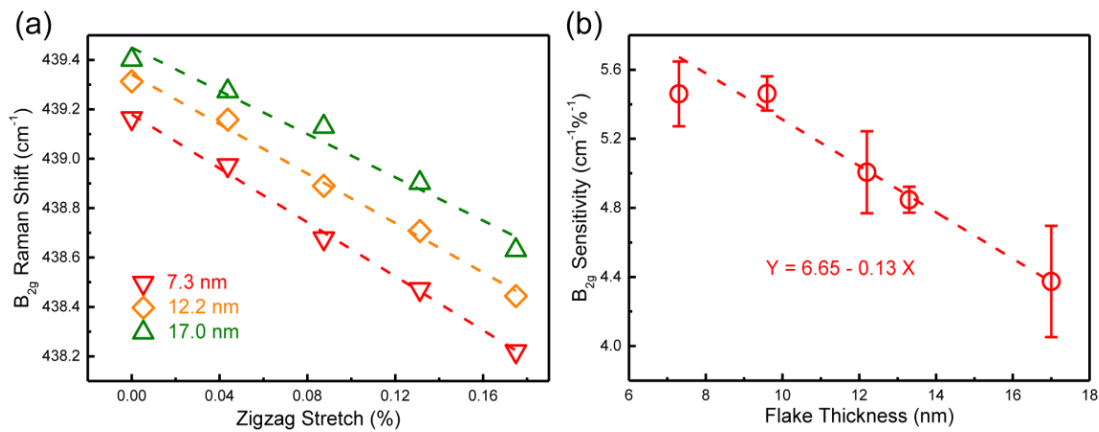

**Supplementary Figure 5 | Layer-dependent studies of uniaxial strained BP.** (a) $B_{2g}$ mode intensity varies with zigzag tensile strains for BP layers with a thickness of 7.3 nm, 12.2 nm, and 17.0 nm. The dashed lines show the linear fitting. (b) Sensitivity of $B_{2g}$ mode changes with different flake thicknesses. The error bar in each thickness is determined from linear fittings in (a). Red dashed line has been given to guide the eyes, where *X* stands for flake thickness in nm, and *Y* is predicted $B_{2g}$ strain sensitivity.

|  | Armchair Strain | |
|---|---|---|
|  | Armchair Polarization (cm$^{-1}$/%) | Zigzag Polarization (cm$^{-1}$/%) |
| $A_g^1$ Stretch | 1.63 | 1.49 |
| $A_g^1$ Compress | 2.10 | 2.09 |
| $A_g^2$ Stretch | No Noticeable Shift | No Noticeable Shift |
| $A_g^2$ Compress | No Noticeable Shift | No Noticeable Shift |
| $B_{2g}$ Stretch | 1.45 | 1.10 |
| $B_{2g}$ Compress | 0.76 | 1.08 |

|  | Zigzag Strain | |
|---|---|---|
|  | Armchair Polarization (cm$^{-1}$/%) | Zigzag Polarization (cm$^{-1}$/%) |
| $A_g^1$ Stretch | 0.61 | 0.4 |
| $A_g^1$ Compress | 0.84 | 0.77 |
| $A_g^2$ Stretch | 0.94 | 1.15 |
| $A_g^2$ Compress | 1.71 | 1.74 |
| $B_{2g}$ Stretch | 1.95 | 1.84 |
| $B_{2g}$ Compress | 1.85 | 1.75 |

**Supplementary Figure 6 | Laser polarization configurations.** Laser polarization configurations with respect to two principle axe strains.

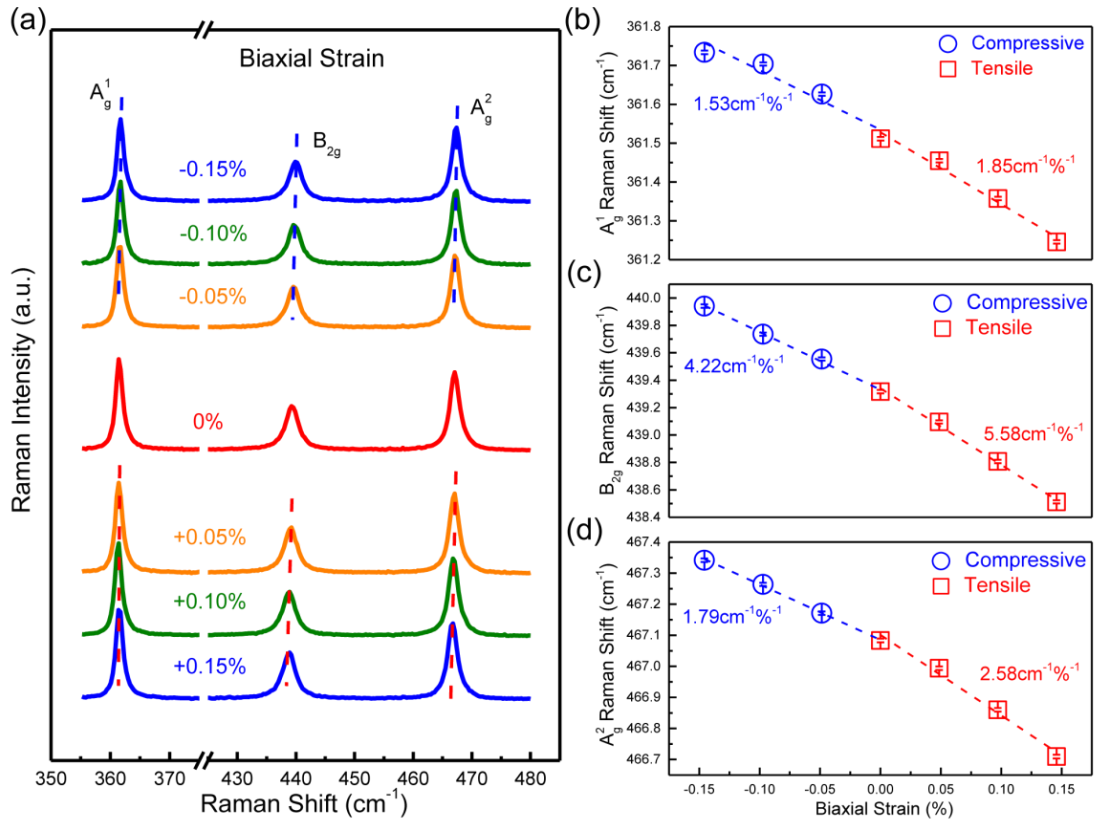

**Supplementary Figure 7 | Raman response in biaxial strained BP.** (a) Raman spectra of BP for both tensile and compressive biaxial strains, as shown in Supplementary Figure 1b with plates for biaxial strains. The dashed lines are here to guide the Raman peak position shift. Raman shifts of (b) $A_g^1$ (c) $B_{2g}$ and (d) $A_g^2$ modes in biaxial strained BP are summarized here. The dashed lines show the linear fittings, and error bars are determined from Lorentzian peak fitting.

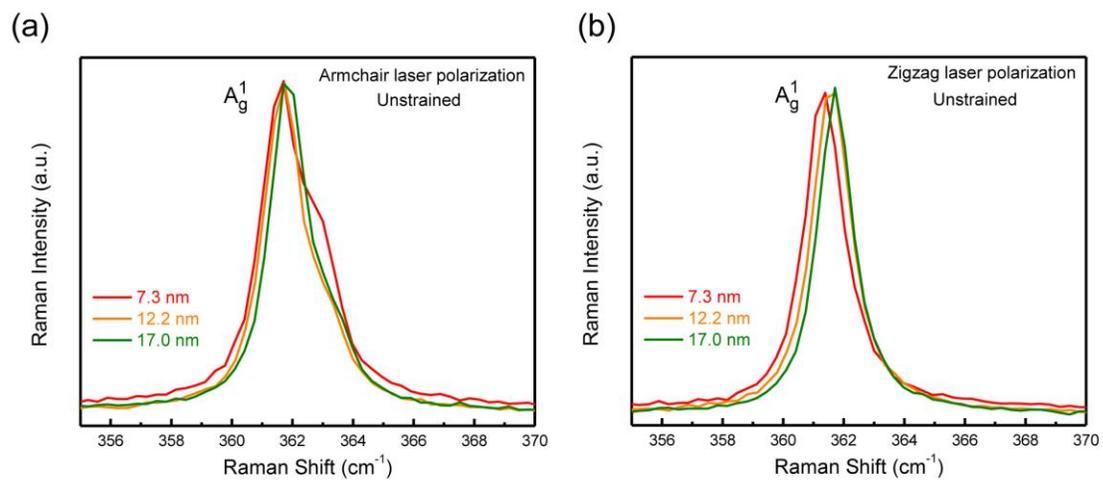

**Supplementary Figure 8 | Shoulder peak in $A_g^1$ mode along different laser polarization directions.** Thickness-dependent $A_g^1$ mode Raman intensity along (a) armchair laser polarization, and (b) zigzag laser polarization.

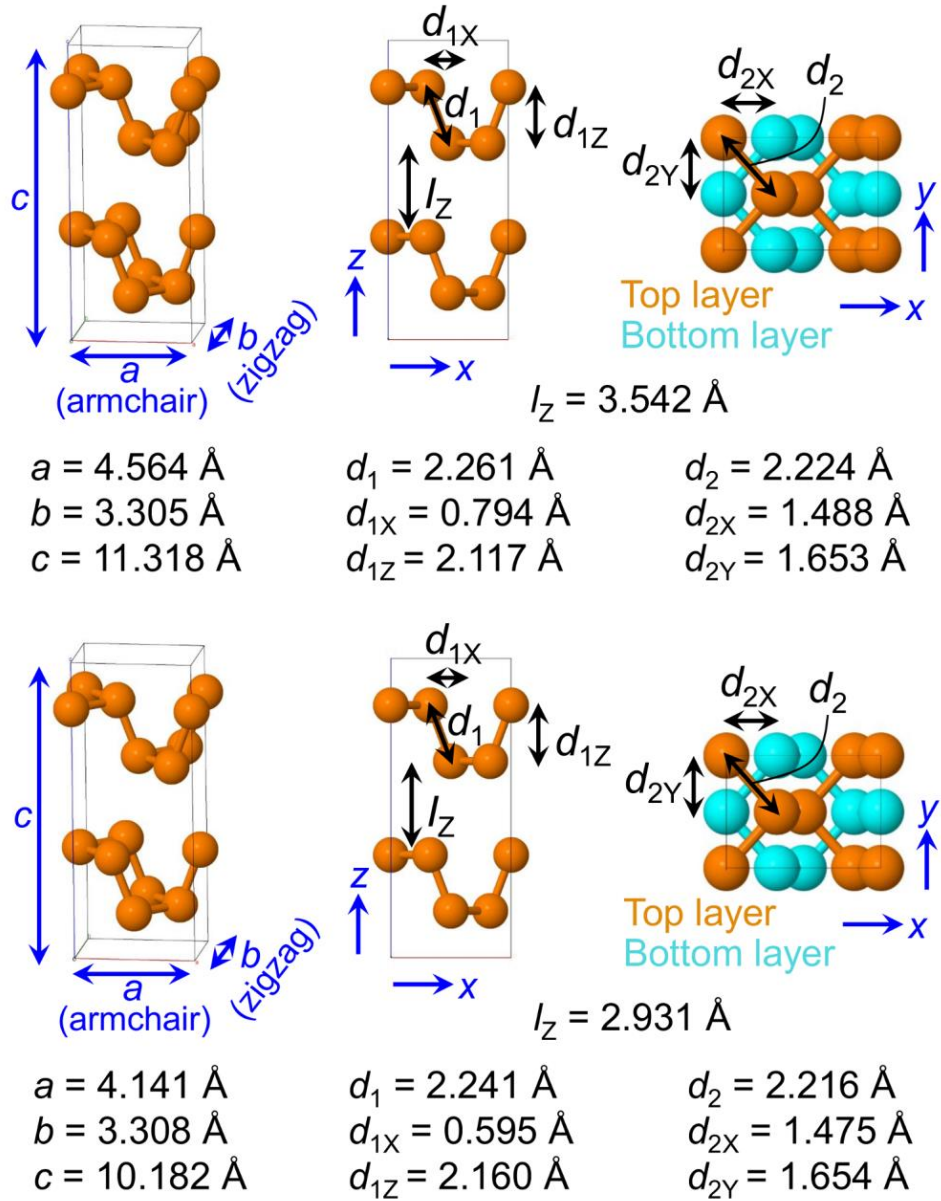

**Supplementary Figure 9 | GGA and LDA structural properties of unstrained BP.** Atomic structure of bulk BP, lattice constants, P-P bonds and their projections along *x*, *y* and *z*. Results obtained from density functional theory using the GGA-PBE (top) and LDA-CA (bottom) approach for exchange-correlation.

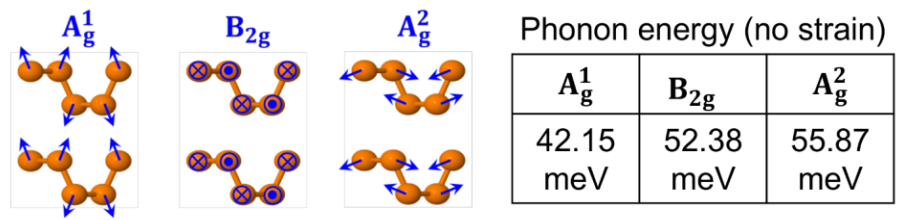

**Supplementary Figure 10 | Calculated GGA phonon energies of BP.** Atomic motion of the prominent Raman-active phonon modes, and their energy (unstrained BP). Results obtained from density functional theory using the GGA-PBE approach for exchange-correlation.

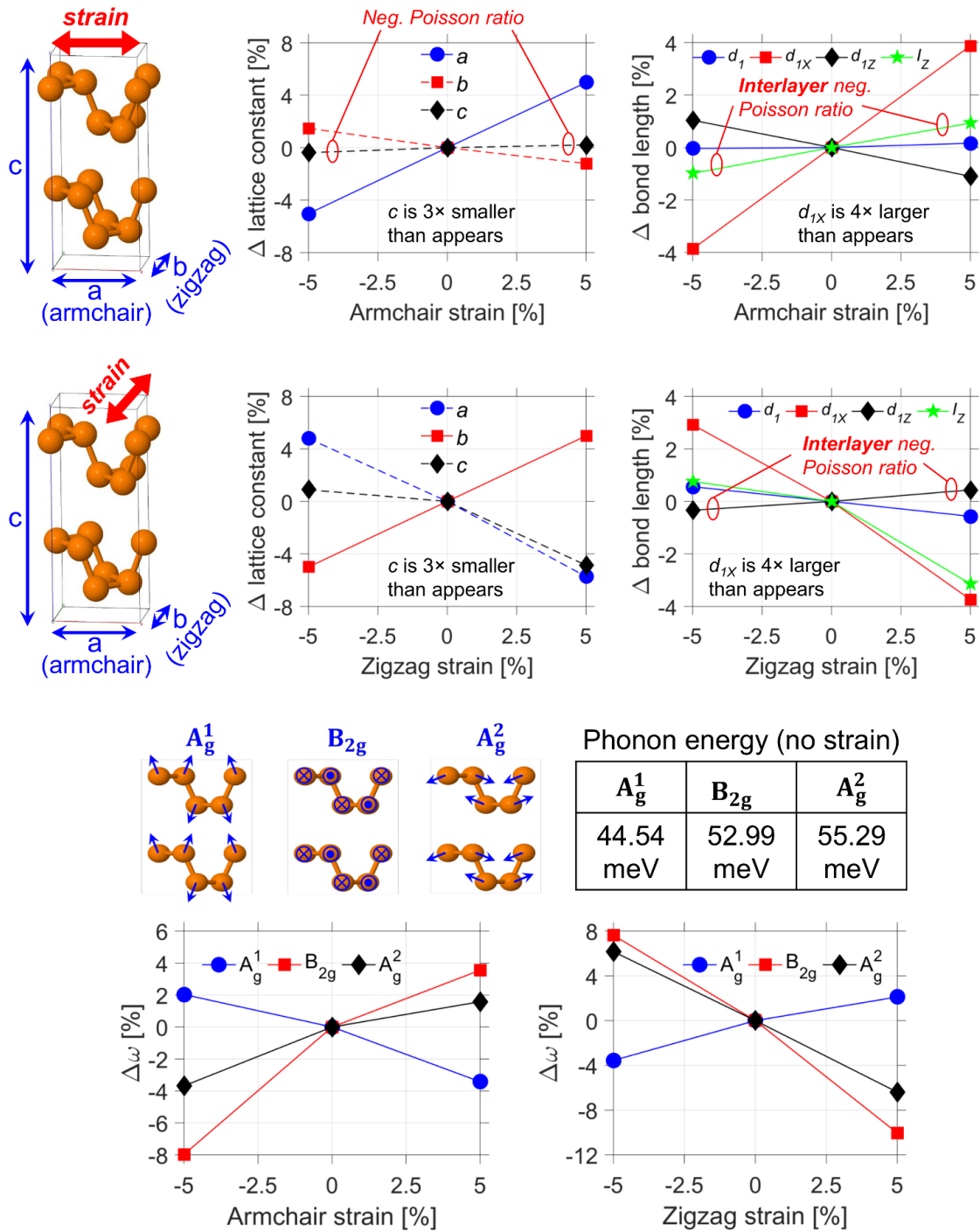

**Supplementary Figure 11 | Calculated LDA structural properties and phonon energy of strained BP.** Variation in lattice constants, bond lengths and their projections versus applied

armchair and zigzag uniaxial strain. Variation in phonon energy with applied armchair and zigzag uniaxial strain. All results obtained from density functional theory using the LDA-CA approach for exchange-correlation.

**Supplementary Note 1 | Uniaxial four-point bending apparatus**

The principle of four-point bending based set-up is well known for its use in mechanical testing of structure materials, and the key idea of using four-point bending is to have a reliable and accurate way of applying a uniaxial strain to samples[1]. For our home-made apparatus design, a uniform strain can be evenly applied, moreover, the value of the applied strain can be accurately controlled. In this section, we discuss the design of our four-point bending set-up in terms of basic operation and mathematical calculation of applied strain. Uniaxial strain both tensile and compressive processes can be achieved by using specific top and bottoms plates, shown in Supplementary Figure 1a. The magnitude of applied stress can be generated accurately by reading the relative displacement of the top and bottom plates through miniature micrometer heads. It should be noted that, our stress bending direction is perpendicular to the Raman polarization laser direction, as shown in Supplementary Figure 1d, indicating that the flake with laser incident along armchair direction would be applied with a zigzag strain, and for the flake with laser incident along zigzag direction would have armchair strain. The fundamental equation describing the magnitude of applied strain is given by[1],

$$\sigma = \frac{3d_Z E t}{a(3L-4a)} \quad (S1)$$

where $\sigma$ is the unit value of stress, $a$ is the length between two nearest supporting points, $L$ is defined as the distance between two farthest supporting points, $t$ points to the thickness of bending wafer, $d_Z$ stands for the deflection plate in vertical direction, and $E$ is Young's modulus. Due to the strong surface adhesion of atomic-layer thin BP flakes on $SiO_2$. The strains applied to Si substrate are transferred to BP flakes.

**Supplementary Note 2 | $A_g^2$ Raman shift in armchair strained BP.**

Along armchair strained BP, $A_g^2$ peak shows an insensitive behavior. As shown in Supplementary Figure 2, we did not obtain a noticeable Raman shift in $A_g^2$ mode with various armchair strains. Raman shift difference between the 0% and ±0.17% strain is less than ~0.1 cm$^{-1}$, and both tensile/compressive strains have not shown any linear shifts. We believe the insensitivity of $A_g^2$ peak can be fairly attributed to the fact that BP structure is anisotropic and much softer along the armchair direction, such that the magnitude of frequency shifts are much smaller than those under zigzag strains.

**Supplementary Note 3 | Multiple loading/unloading processes**

To evaluate the strain effect, we conducted fundamental calibration measurements of multiple loading/unloading processes of a BP film, regarding to the specific $B_{2g}$ mode. Supplementary Figure 3 shows the Raman spectra of $B_{2g}$ frequency at five conditions: unstrained, $1^{st}$ +0.13% strain, released, $2^{nd}$ +0.13% strain, and $2^{nd}$ released. The red shift of the active mode under strain as well as the blue shift back to original position due to strain relaxation are clearly observed in our experiment, indicating a good strain reversibility of BP. In order to further verify the quantity of strain effect for multiple cycles, we computed the peak frequency of $B_{2g}$ mode by generating Lorentzian curves to fit the spectra. The peak frequency of $B_{2g}$ before any application of strain locates at 439.40 cm$^{-1}$. We noticed that the $B_{2g}$ peak position here is a little bit larger than the one in main text, and it is attributed to the variation of flake thickness[2], which will be discussed later in the Supplementary note 5. The first +0.13% tensile strain has red-shifted the peak to 439.10 cm$^{-1}$, with a step of 0.30 cm$^{-1}$. Once the strain is released for the first time, the $B_{2g}$ peak immediately goes back to 439.29 cm$^{-1}$. For the second +0.13% tensile strain, the magnitude of active mode peak signal has been received at 439.02 cm$^{-1}$, with a step of 0.27 cm$^{-1}$. In our two cycles, uniaxial strain has demonstrated a similar quantity of red-shift, revealing a high repeatability of our experiment and a great elasticity of BP film. It is important to mention that a slight offset during the unloading process is observed in our research. In order to validate all our experiment, uniaxial strain has been gradually applied from 0% up to ±0.17% monotonously, and then been released to start another strain direction. Under such circumstance, the minor effect of unloading offset can be fairly neglected in the analysis.

**Supplementary Note 4 | Air-stability of Raman spectra**

We conducted our BP Raman experiments in ambient atmosphere without special passivation. Towards to the less stability of thin BP film, we examine the air-stability of Raman spectra of BP in ambient environment through a comprehensive suite of time dependent measurements. The exfoliated BP flake has been stored, and measured repeatedly in ambient. Air-stability of Raman spectra has been monitored through measuring the Raman spectra as a function of time. As shown in Supplementary Figure 4a, the Raman spectra of thin-film BP demonstrates a quite stable characteristic within a measured time frame. For each vibration mode, the peak position does not provide a systematical shift, which indicates Raman shift of uniaxially strained BP is directly associated with the strain induced interatomic vibration, rather than the potential degradation of BP film. In a normal case, it took approximately 2 hours to perform a set of strain experiments for a single flake. As reflected from Supplementary Figure 4b, the deviations from three modes within 10-hours-time period is extremely small as compared to strain induced Raman frequency shift, and can be fairly ignored in strain analysis.

**Supplementary Note 5 | Layer-dependent Raman spectra of uniaxial strained BP**

Layer-dependent Raman spectra of uniaxial strained BP has also been performed in our experiment. BP films with a thickness of 7.3 nm, 9.6 nm, 12.2 nm, 13.3 nm, and 17.0 nm have been examined through a comprehensive analysis of uniaxial tensile and compressive processes. All the flakes present the same Raman spectra transitions, indicating a high repeatability of our experiment and a solid conclusion of experimental results. $B_{2g}$ mode in zigzag strain with the largest sensitivity has been selected to demonstrate the layer-dependent behavior. As shown in Supplementary Figure 5a, three samples all illustrate a monotonic variation with applied strains, demonstrating a clearly red-shift in stretched zigzag direction. Interestingly, we notice that there is an obvious frequency shift on $B_{2g}$ mode as a function of thickness at initial condition without any strains, where the peak position moves towards lower Raman shift by a magnitude of 0.24 cm$^{-1}$. The classical model for coupled harmonic oscillators can be used to explain this effect as the vibrations of bulk materials built up from van der Waals bonded layers, which predicts a stiffening process in $B_{2g}$ mode with the additional layers adding to form the bulk sample from few layers[3]. We should mention that, laser scanning from Raman system may be accompanied by a layer-thinning procedure, however, coupled harmonic oscillator induced Raman shift per nm is extremely small (0.025 cm$^{-1}$nm$^{-1}$) as compared to the strain-induced signal, which can be neglected in our analysis. Meanwhile, strain sensitivity derived from the slope of Raman shift versus strains has been determined and plotted in Supplementary Figure 5b for all five flakes with different thicknesses. Comparing with thick samples, thinner BP flake has presented a pronounced strain sensitivity, which is in line with previous observation in graphene strain experiment[2], as well as previous layer dependent Raman studies with respect to BP[3,4]. In addition, we have tried to fit the sensitivity behavior in a linear line, and we end up with an experimental prediction of strain sensitivity varies with different flake thickness.

**Supplementary Note 6 | Laser polarization configurations.**

For the results shown in main manuscript, the strain direction is perpendicular to the laser polarization direction based on our apparatus set-up (Supplementary Note 1), *i.e.* armchair strain is recorded by performing zigzag Raman polarization, and zigzag strain is recorded by performing armchair Raman polarization. To further confirm the strain sensitivities of BP, we have conducted two separated experiments to demonstrate laser polarization does not have influence among three dominate Raman peak under principle strains. As depicted in Supplementary Figure 6, armchair and zigzag strain sensitivities have been reported by utilizing two distinct laser polarizations, ending up with a total of four cases. Summarized strain sensitivities indicate the same shift characteristic and similar magnitude under different laser polarizations. In conclusion, we have excluded polarized laser direction induced strain effect in all our experiments.

**Supplementary Note 7 | Raman response in biaxial strained BP.**

Few-layer BP has been exfoliated from the bulk crystal by standard scotch tape method and transferred to a conducting Si substrate with a 300 nm $SiO_2$ capping layer. The biaxial strain has been applied on the flake for both tensile and compressive strains (see Supplementary note 1), and Raman evolution of biaxial strain has been summarized in Supplementary Figure 6. All three dominated peaks are red-shifted under tensile strain, and blue-shifted under compressive strain. The magnitudes of strain sensitivity are labeled on Supplementary Figures 7b-d.

**Supplementary Note 8 | Shoulder peak in $A_g^1$ mode along different laser polarization directions.**

The dominate $A_g^1$ mode peak overlaps with another peak appearing as a weak shoulder in Figure 4 in main manuscript. This phenomenon only happens along the armchair laser polarization, and it has been observed in our previous studies as well[5]. To indicate that the weak shoulder in $A_g^1$ peak is directly associated with Raman laser polarization direction, we then plot the Raman intensity for both armchair and zigzag laser polarization with respect to different flake thicknesses. Supplementary Figure 8a indicates Raman spectra along armchair laser polarization. Flakes with different thicknesses have all shown a weak shoulder in $A_g^1$ mode. However, as we switch the laser to be zigzag polarization, depicted in Figure 8b, the weak shoulder disappears. The results here have demonstrated that the splitting peak shoulder in $A_g^1$ mode is related to anisotropic Raman laser polarization. It might be related to the second order Raman processes[6]. Meanwhile, at unstrained situation, the $A_g^1$ peak position is presenting a blue-shift as flake thickness increases, which is attributed to the layer-dependent BP Raman spectra (see Supplementary note 5) and our results in $A_g^1$ mode are consistent with Supplementary note 5 and previous work[4].

**Supplementary Note 9 | GGA and LDA structural properties of unstrained BP.**

Both GGA and LDA models have been introduced to perform theoretical simulations. Atomic structure of bulk BP, lattice constants, P-P bonds and their projections along $x$, $y$ and $z$ are listed here for both GGA and LDA models.

**Supplementary Note 10 | Calculated GGA phonon energies of BP.**

Phonon energy at unstrained situation for $A_g^1$, $B_{2g}$ and $A_g^2$ are listed at top. Meanwhile, variation in phonon energy with applied armchair and zigzag uniaxial strain are shown here for all three main peaks.

**Supplementary Note 11 | Calculated LDA structural properties and phonon energy of strained BP.**

Variation in lattice constants, bond lengths and their projections versus applied armchair and zigzag uniaxial strains. Results obtained from DFT using the LDA-CA approach for exchange-correlation. Atomic motion of the prominent Raman-active phonon modes, and their energy (unstrained BP). Variation in phonon energy with applied armchair and zigzag uniaxial strain. Results obtained from DFT using the LDA-CA approach for exchange-correlation are consistent with using the GGA-PBE approach for exchange-correlation, as presented in main manuscript.